# Replacing P values with frequentist posterior probabilities - as possible parameter values must have uniform base-rate prior probabilities by definition in a random sampling model

Huw Llewelyn

Department of Mathematics

Aberystwyth University

Penglais

Aberystwyth

hul2@aber.ac.uk

## Abstract

Possible parameter values in a random sampling model are shown by definition to have uniform base-rate prior probabilities. This allows a frequentist posterior probability distribution to be calculated for such possible parameter values conditional solely on actual study observations. If the likelihood probability distribution of a random selection is modelled with a symmetrical continuous function then the frequentist posterior probability of something equal to or more extreme than the null hypothesis will be equal to the P-value; otherwise the P value would be an approximation. An 'idealistic' probability of replication based on an assumption of perfect study methodological reproducibility can be used as the upper bound of a 'realistic' probability of replication that may be affected by various confounding factors. Bayesian distributions can be combined with these frequentist distributions. The 'idealistic' frequentist posterior probability of replication may be easier than the P-value for non-statisticians to understand and to interpret.

## 1. Introduction

It is well recognised that if the prior probabilities of possible parameter values during random sampling are assumed to be uniform (or 'non-informative' or 'flat') then according to Bayes rule, the probability of the null hypothesis or something more extreme will equal the P-value [1, 2]. It was on this basis that Bayes [3] calculated the probability of a true proportion falling within any specified range during random sampling by applying the binomial distribution. In order to do this he appeared to use Bayes rule and to assume uniform priors (but did not state either explicitly). I shall show here that Bayes was correct to think that during random sampling the postulated parameter values can be modelled using equal 'base-rate' prior probabilities. It is thus possible to apply Bayes rule in a way that may make statistical inference easier to understand by non-statisticians, an issue that is of concern at present [4] and which may be contributing to the so-called 'replication crisis'.

## 2. A source of confusion

Confusion as arisen by regarding a population parameter $pi$ (e.g. the proportion of bilingual people in a sub-population) as one of a series of sub-population parameters $p1, p2, \dots pm$ as shown in figure 1. If we wish to calculate the proportion of bilinguals who belong to the 48% sub-population for example, we can do so with Bayes rule. The proportion of the population that falls into the '48% sub-population' is 0.073 and since 48% of these are bilingual, then the proportion of the total population that is in this 48% subgroup and also bilingual is 0.073*0.480 = 0.038. However, the proportion that is bilingual in the total population is 0.475, so the proportion of those who are bilingual who are in the 48% sub-population by Bayes rule is 0.073*0.480/0.475 = 0.080. The problem of estimating the value of a parameter is quite different to this however.

**Figure 1. Distribution of sub-population parameters for true proportions of those who are bilingual**

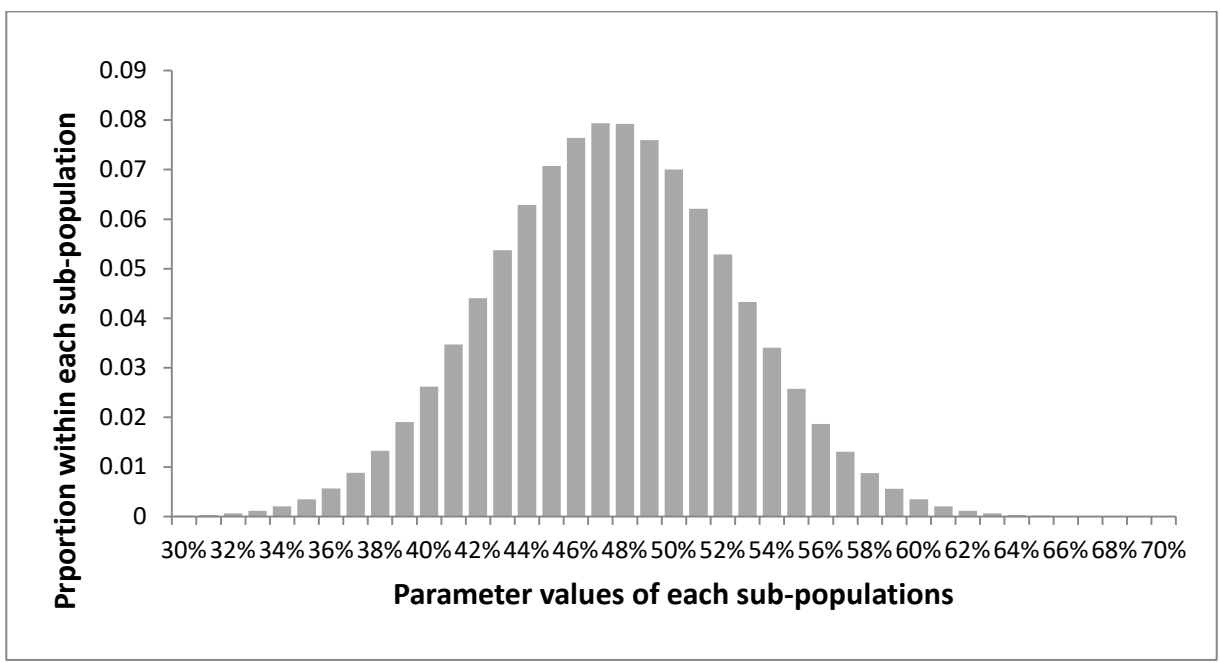

If we are presented with one of the sub-populations from 0% to 100% but not told which it is, and are asked to estimate its parameter value $pz$ by taking random samples from that unknown sub-population, the result would depend only on the proportion of bilingual people in the sample. The relative sizes of the sub-populations would not affect the result. As the sample size increases, the sample proportion converges on the 'true' value $pz$ equally rapidly whether it turned out to be a large more common sub-population (e.g. the '48%' sub-population) or a smaller rarer sub-population (e.g. 'the 32%'). Therefore the 'prior probability' of the sub-population in the source population is not relevant to the result of a sampling process and to include it inappropriately would bias the random sampling process.

A very large sample would converge on the 'true' value of $pz$ in the sub-population. We can therefore consider a series of hypothetical results based on such a very large sample. These can be regarded as possible 'model' subsets, each one containing the same number of elements (i.e. the same large number of random selection results). This equal number of elements in each subset means that the prior probability of each possible 'true' result subset would be the same (i.e. the priors are uniform). We can choose any number 'm' of 'possible outcome subsets' for our model depending on the precision required when estimating the value $pz$. If we then make 'n' selections from each of these 'possible outcome-subsets' then the result of these n selections can be represented by X1, X2, … Xn, any one of these being Xj.

## 3. Random sampling model

A random sampling model is represented by:

$$f(pi|Xj) = n/m . f(Xj|pi)$$

when $pi$ is any one of possible parameter values $p1, p2, … pm$ corresponding to the 'm' hypothetical sub-sets $\{p1\}, \{p2\}, … \{pm\}$. Each hypothetical subset $\{pi\}$ is made up of a large but equal number of elements, each element being a single random selection from a population with an unknown parameter $pz$ and are therefore equally probable so that the 'base rate' prior probability of any element having the attribute $pi$ is $f(pi) = 1/m$. $Xj$ represents any result of n random selections from a hypothetical population $\{pi\}$, the possible outcomes of $Xj$ (when j = 1 to n) being $X1, X2, … Xn$. Because any set $\{Xj\}$ is based on an equal number of 'n' elements (i.e. n individual selections), then the 'base rate' prior probability of any element within $\{Xj\}$ is $f(Xj) = 1/n$.

When n=m, $pk$ is a particular $pi$ and $Xy$ is a particular $Xj$ then $f(pk|Xy) = f(Xy|pk)$. When $f(Xj \leq Xy|pk)$ represents a 'P-value' and when $f(Xj \leq Xy|pk) = f(pi \geq pk|Xy)$, the latter is a frequentist posterior probability for the range $\{pi \geq pk\}$.

Let Xj' be any X1', X2', … Xn' and Xj'' be any X1", X2", … Xn" (when n' + n" = n). Let the function 'g' combine the likelihood distributions for Xy' and Xy'' (the latter being particular values of Xj' and Xj") to give the joint likelihood distribution of:

$$g\big(f(Xy'|pk), f(Xy''|pk)\big) = f(Xy'^{\wedge}Xy''|pk) = f(Xy|pk).$$

The posterior odds of $\frac{f(pk|Xy')}{f(\sim pk|Xy')}$ can be regarded as new non-base rate prior odds to be updated by a second random sampling result of $Xy'$ to give the posterior odds of:

$$\frac{f(pk|Xy'\wedge Xy'')}{f(\sim pk|Xy'\wedge Xy'')} = \frac{f(pk|Xy')}{f(\sim pk|Xy')} * \frac{1/m * f(Xy''|pk)}{1/m * \sum_{i\neq k}^{m} f(Xy''|pi)} = \frac{f(pk|Xy)}{f(\sim pk|Xy)}$$

So that $f(pk|Xy'\wedge Xy'') = f(pk|Xy)$ is a second 'posterior probability'.

The above prior probabilities $f(Xj) = f(Xj|U)$ and $f(pi) = f(pi|U)$ are 'base rate' prior probabilities because $\{U\}$ represents a universal set of the possible outcomes $\{pi\}$ and $\{Xj\}$ of random selections such that $\{Xj'\} \subseteq \{U\}$, $\{Xj''\} \subseteq \{U\}$, and $\{Xj\} \subseteq \{U\}$. The 'non-base rate' prior probability $f(pk|Xy')$ is not a 'base rate' prior probability with respect to the likelihood $f(Xy''|pk)$ because $\{Xj'\}$ is not a universal set and so $\{Xj''\} \nsubseteq \{Xj'\}$.

## 4. An example: Plotting distributions

We have a set of the language records of a population with an unknown proportion (i.e. '$pz$' in section 3) of bilingual people. We choose a range of possible true proportions of those who are bilingual in that unknown population, this range of true proportions $pi$ being from $p1=0\%$ to $p100=99\%$ (these corresponding to $p1\ to\ pm$ in section 3). We then create 100 model subsets that contain 100 possible 'true' values of $pz$ from 0% to 99%. Each of these 100 model subsets would contain the same ideally large number of elements (i.e. model sample record cards) so that there will be the same number of elements in each set. Therefore the prior probability of each model subset representing the unknown proportion would be 1/100 = 0.01 (i.e. '1/m' in section 3).

If we took a sample from the real source population $\{pz\}$ and found a proportion of 50/99, we can then calculate the likelihood of selecting 50/99 from each of the 100 model subsets. For example, the likelihood of selecting 50/99 from 'model subset 43%' using the binomial distribution would be 0.02595 and from 'model subset 59%' it would be 0.01869. As there are 100 possible results from 99 selections, the prior probability of each possible result from 0/99 to 99/99 would also be 0.01, so that the prior probability (i.e. $f(Xj)$ in section 3) of having selected 50/99 would be 1/n = 1/100 = 0.01. This means that the posterior probability of selecting the possible parameter of 'represented by model subset 43%' (i.e. $f(pk|Xy)$ in section 3) after observing a sample of 50/99 would be equal to the likelihood of selecting 50/99 from a population with a true proportion of 43% (i.e. $f(Xy|pk)$ in section 3) The posterior probability that the sample 50/99 came from 'model sub-set 43%' is therefore:

$$1/(1 + 0.01/0.01 * 0.01869/0.02595) = 0.58118.$$

## 5. Comparing likelihood and binomial distributions

When the prior probability of the parameter and statistic are the same, the likelihood distribution for 50/99 is the same as the posterior probability distribution of each possible true value conditional on the randomly selected sample of 50/99. This is shown by the solid line in figure 2. Furthermore, this posterior distribution should be the same as a binomial distribution created by making 99 selections (with possible 100 outcomes) from a population with a proportion of 50/99 = 0.50505. This binomial probability distribution is represented by the round solid markers in figure 2. The binomial and the normalised posterior distributions are indeed superimposed when calculated with an Excel spreadsheet and a plot of each normalised likelihood probability against the binomial probability is a line of identity.

**Figure 2: Comparison of distributions of frequentist posterior probabilities of each possible 'true' proportion conditional on an observed result of 55/99 and a binomial distribution when P = 55/99 = 0.505**

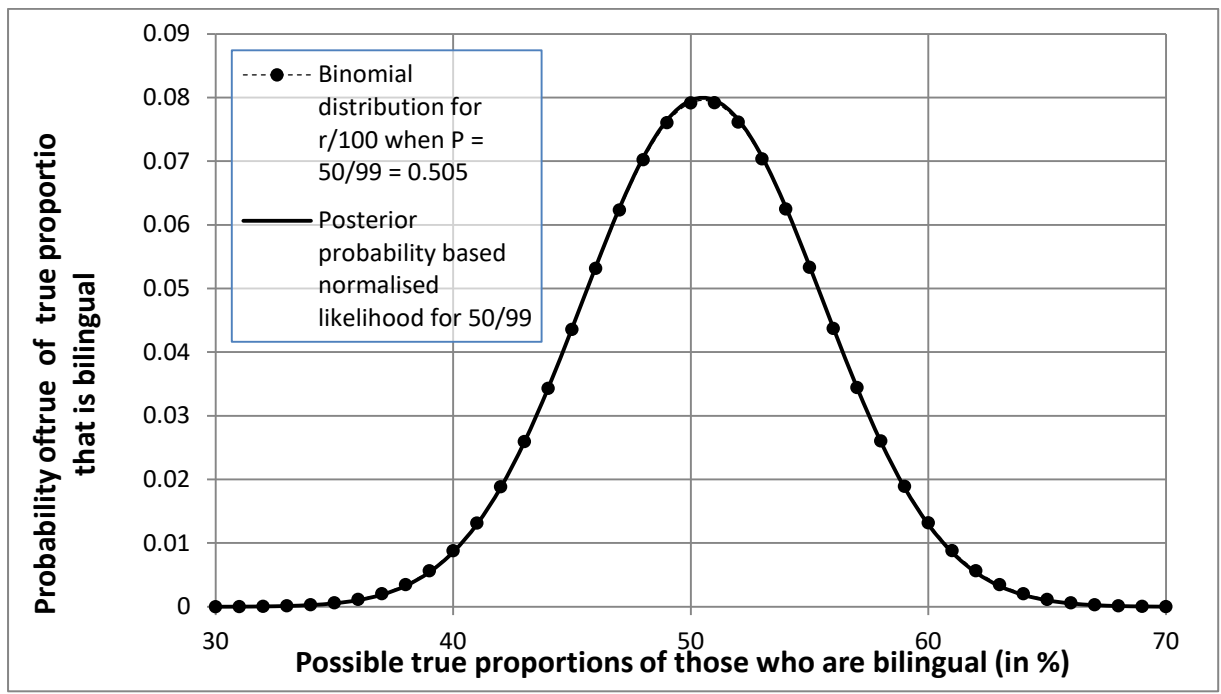

In order that the sum of the likelihood probabilities created by the binomial distribution from 0/99 to 99/99 sum to 1, the number of 'true' data points has to be 101 (i.e. 0%, 1%, … 99% and 100%). However, this means that there would be 100 'spaces' between the 101 'true' data points, so that it is the prior probability of the 100 'spaces' that are equal to the prior probabilities of the 100 possible observations from 0/99 to 99/99. Each of these 'spaces' will be represented by the pair of values that form its upper and lower bounds. Interestingly, this corresponds to the idea of 'induced probability pairs' described elsewhere [5]. This involves regarding the probability to be applied to a new 100th member of a set of 50/99 as being equal to 50/100 if the new member does not have the predicted attribute (e.g. not being bilingual) or 51/100 if the new member does have the attribute (e.g. of being bilingual). So, the future induced probability pair based on a past experience of 50/99 would be '0.50, 0.51'. The 101 data points in figure 2 can therefore be regarded as a chain of 100 induced probability pairs that arise from the observed proportions from 0/99 to 99/99.

### 6. Specifying and comparing possible 'true proportions

Instead of the 101 possible 'true' proportions in figure 2, if we had specified 10, 001 data points (i.e. 0.00%, 0.01%, 0.2% up to 99.99%, 100.00%), then the 'prior probability' of each possible 'true' proportion would have been 1/10,000 = 0.0001 instead of 0.01. The likelihood probability of selecting 50/99 from 43.00% would have been 0.02595. However, the posterior probability of seeing 43.00% (with a prior probability of 0.0001 as opposed to the prior probability of 43% of 0.01) would not be equal to the corresponding likelihood but 1/100$^{th}$ of it: 0.0002595. This is because the prior probability of selecting 50/99 is 0.01 and the prior probability of a 'true' result of 43.0% is 0.0001, so the posterior probability of 43.00% is 0.01*0.02595 / 0.0001 = 0.0002595.

### 7. Normalisation

If the likelihood probabilities of randomly selecting 50/99 were added for each of the possible 'true' proportions from 0.00% to 100.00% they would sum to 100. The process of dividing each of the likelihood probabilities by their sum (of 100 in this case) is called 'normalisation'. This will provide the posterior probability of each possible 'true' result. It is the same calculation as that performed in the previous paragraph.

### 8. Combining probability distributions using Bayes rule

Figure 3 shows three distributions. The dotted line shows a 'prior probability' distribution obtained by 'normalising' the likelihood distribution for 22/46. The solid line represents the normalised likelihood distribution for 28/53 and the double line the posterior distribution obtained by multiplying the prior probabilities for the 'broken line' 22/46 distribution by those of the solid line 28/53 distribution and then 'normalising' the results. The double line posterior distribution can also be obtained by 'normalising' the likelihood distribution for (22+28)/(46+53) = 50/99. Note that the height of the double line distribution is 1/100ths of the height of the same distribution in figure 2. This is because in figure 2, the prior probability of each possibility was 0.01 whereas in figure 3, the prior probability of each possibility was 0.0001 so the posterior probabilities of each possible result in figure 2 were 100 times those in figure 3.

**Figure 3: Prior, 'normalised likelihood' and posterior probabilities of the possible true proportions who are bilingual**

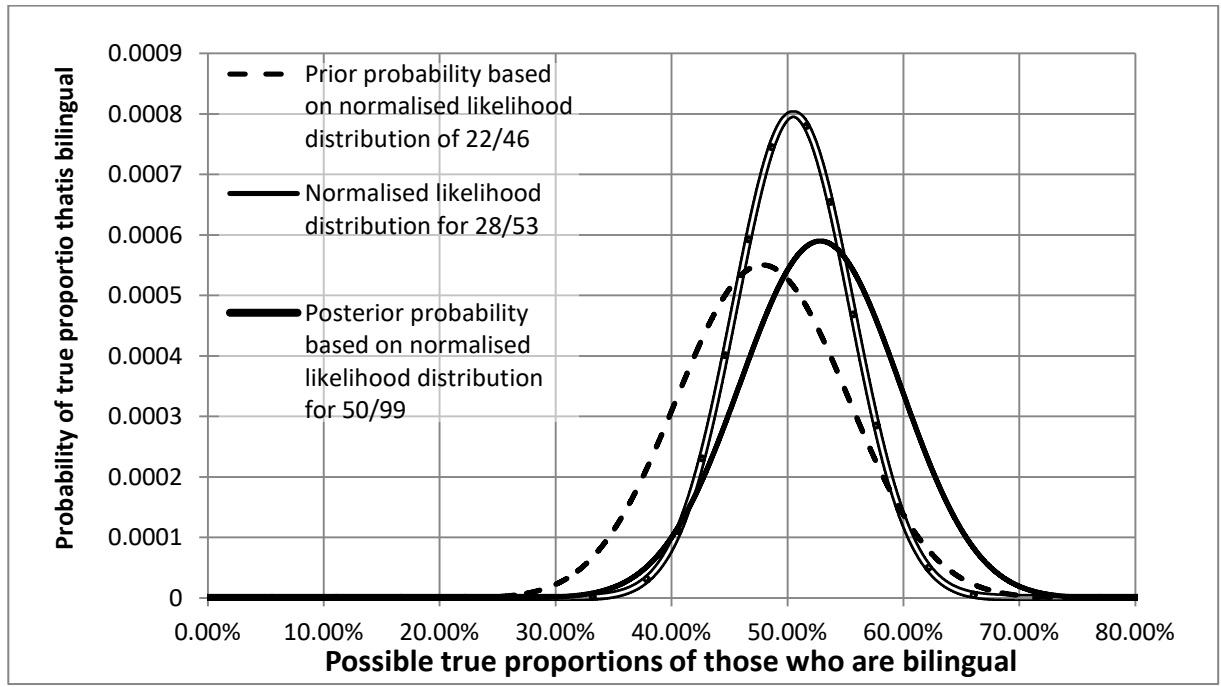

The number of possible true outcomes to be considered when calculating normalised probabilities affects the precision of calculations. It can be seen by comparing figures 2 and 3 that the shape of the distribution is the same and that the probabilities of replication within a fixed range is approximately the same, the slight difference being due to the increased precision of the calculation when more possible outcomes (e.g. 10,000 in figure 3 compared to 100 in figure 2) are used in the calculation.

The number of outcomes models the size of future studies during the replication process. For example, figure 2 shows the distribution of possible outcomes of repeating studies with 100 possible observations per study whereas figure 3 shows the distribution when 10,000 observations are included in each study. In both cases the individual studies (containing 100 or 10,000 observations) are repeated an 'infinite' number of times, the estimates of outcome in both cases being based on the observation of 50/99 in the original study.

If the likelihood distribution is modelled with a continuous function then the posterior probabilities would approach zero. To make sense they would be have to be integrated to provide cumulative probabilities so that the probability of a 'true' result falling into a range can be calculated by subtracting one cumulative probability from another.

## 9. P-values and posterior probabilities

In figure 4, there are two distributions. The solid line is a Gaussian distribution calculated by using 10,001 data points, each therefore with a prior probability of 0.0001 compared to a prior probability of 0.01 for the observation of 50/99 so that the posterior probabilities are again all 1/100$^{th}$ of the value of those in figure 2 but the same as those in figure 3. The broken line is a Gaussian distribution with 10,001 data points modelling the binomial distribution for 100 possible outcomes based on a probability of 0.404 = 40/99.

**Figure 4: Gaussian estimate of the normalised binomial likelihood distribution for 50/99 and the binomial distribution for r/100 when null hypothesis is 40.4%**

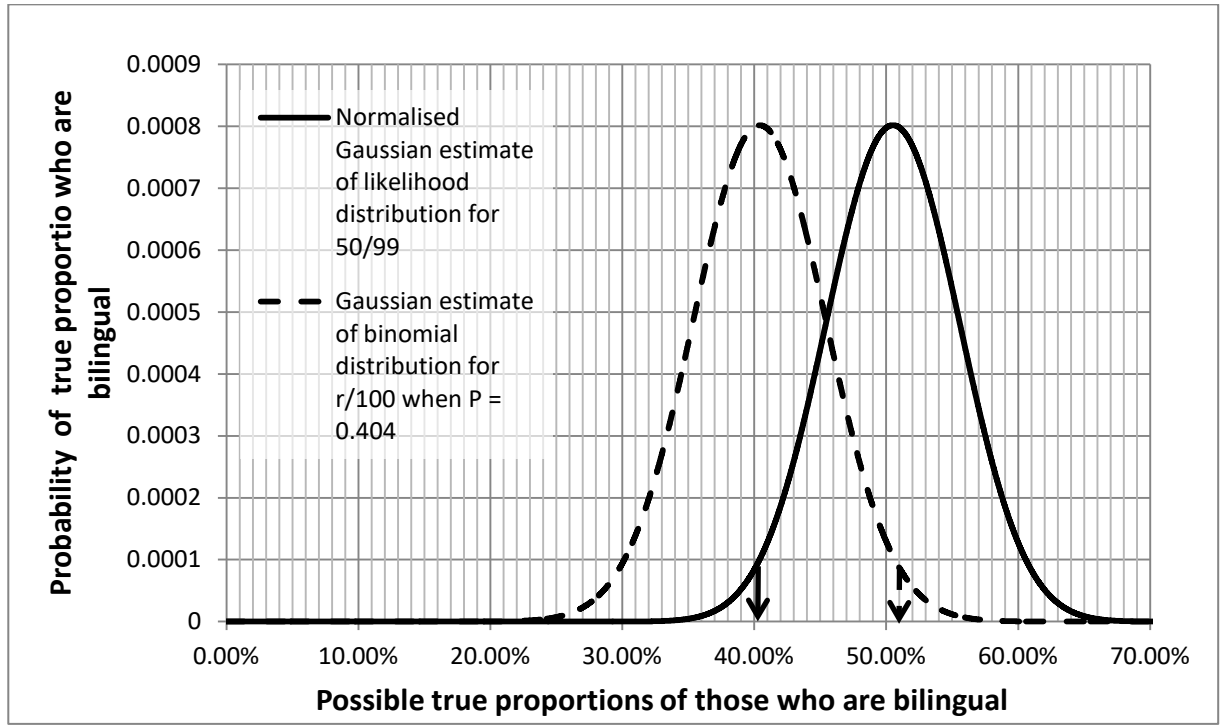

A line drawn down from the mode of the broken line probability distribution marks a tail area on the solid normalised likelihood distribution (i.e. 40.40% or lower). A line drawn down from the mode of the solid normalised likelihood distribution marks the tail of the dotted probability distribution (i.e. 50.5% or higher). These tail areas are approximately the same. If we chose 40.40% as our 'null hypothesis' then likelihood probability of 50.50% or more (i.e. more extreme) would be 0.0225, which is the P value. By contrast, the posterior probability of the null hypothesis of 40.4% or less is 0.0194. So the 'P-value' of 0.0225 would only provide an approximation to this frequentist posterior probability of 0.0194. The frequentist posterior probability of value greater than 40.40% would be 1-0.0193 = 0.9807, the approximate value from the P value being 1 - 0.0225 = 0.9775.

## 10. True outcome ranges and replication ranges

10.1 Calculating the probability of a result falling into a range

The probability that the 'true outcome' lies within any range (e.g. 40% to 60% or above 40%) can be calculated by adding all the discrete probabilities within that range. (In the case of continuous distributions, this is achieved by integrating the likelihood probability densities and normalising them.) The range can have an upper and lower bound (e.g. 40% to 60%) in an analogous way to a credibility or confidence interval. Alternatively, the range can be dichotomous by analogy with a null hypothesis (e.g. above or below 40%). They could be chosen by specifying 95% posterior probability ranges with a probability of 0.025 that a true interval would fall into each tail. These ranges would summarise the data and provide an 'idealistic' probability of replication, which means that all observations made during the initial study were assumed to be made with impeccable accuracy and consistency and any subsequent studies were also done in an identical way to the original study so that they could be modelled by the mathematics of random selection.

### 10.2 The choice of replication range

A scientist could specify a replication range which would satisfy colleagues that a repeat study had produced a result similar enough to the original for purposes of scientific hypothesis testing or practical application e.g. for treating patients. If a study was performed and described impeccably in the same setting and could be repeated equally impeccably, then the probability of a future result falling into the 'replication range' would be accurate. Instead of specifying a replication range, the range could be adjusted so that there was a frequentist posterior probability of 0.95 or 0.99 (for example) of a study being replicated within that range. It would be analogous to a 95% or 99% credibility or confidence interval.

## 11. Discussion

### 11.1 Symmetrical and skewed probability distributions

In figure 2, the normalised binomial likelihood distribution and the binomial probability distribution are superimposed so that when one is plotted against the other they form a line of identity. Their modes are both near 0.5 (i.e. 50/99) and therefore almost symmetrical. However, if the 'study result' was nearer 1 (e.g. 95/99) or 0 (e.g. 5/99) then both distributions would be skewed. When plotted using Excel they are not completely superimposed and the plot is not a line of identity but a narrow ellipse around a line of identity with an average gradient of 1. This may be due to calculation errors in Excel but would be interesting to investigate further.

The Gaussian representation of a binomial likelihood distributions also tends to become skewed if its mode is greater or less than a probability of 0.5, this increasing as it approaches 0 and 1. However, the Gaussian representation of the binomial distribution with its mode at the null hypothesis remains symmetrical when the mode is very near to 1 or 0 and would overlap 0 or 1 inappropriately. This means that the correspondence between 'P values' estimated from the symmetrical distribution around the null hypothesis and a skewed probability distribution of the normalized likelihood tail beyond the null hypothesis is only approximate for proportions.

The likelihood distributions for 'true means' of measurements (as opposed to proportions) may be symmetrical, so the 'P value' based on a Gaussian or a 't-distribution' would correspond to the posterior probability of a 'replication range' based on the null hypothesis or something more extreme. It is important of course for non-statisticians to appreciate that the accuracy of these probabilities depends on how well the 'mathematical model' represents 'reality' and that the result is inevitably an 'estimate' even though it may be calculated to many significant figures.

## 11.2 The instability of probability thresholds

If a study that had provided a probability of replication of 0.95 based on an observation of 50/99 was repeated a very large number of times, its mean proportion would be 50.5% on average but therefore less than 50.5% about half the time and more than 50.5% about the other half of the time. However, if we regarded P = 0.05 (equivalent to an 'idealistic' probability of replication of 0.95) as a critical cut-off point for 'statistical significance', then about 50% of repeat studies would not be 'statistically significant'. This sounds alarming, when in reality it is due to unremarkable 'noise'. Goodman [6] has pointed out that this would be the case if we could 'assume' equal prior probabilities of the long term outcomes of random sampling. However, it has been shown here that it is not necessary to 'assume' such equal prior probabilities; this is a special property of random sampling by definition. It may be helpful therefore to replace two-sided P-values (e.g. 0.05) with the corresponding 'idealistic' probability of replication (e.g. 0.95). This might avoid the mistake of thinking that P more than 0.05 (e.g. 0.06) was 'not significant'. It would be clear that an 'idealistic' probability of replication of 0.95 is hardly different from that of 0.94 even though 0.95 corresponds to the concept of 'being statistically significant' and the 0.94 does not.

## 11.3 The width of replication ranges

Another difficulty is that basing a replication range on a null hypothesis of no difference at all may be too flattering because of its width. For example, if we are comparing two treatments and wished to know if the new treatment was better than an old treatment that had a 'cure' rate of 40.4%, then a null hypothesis of 40.4% response would include bare differences (e.g. of 40.5%) in the replication range. This difference of 40.5 – 40.4% = 0.1% represented by a new treatment response rate of 40.5% may not be helpful in practice. It would be up to the individual who wished to act on the information to decide how much better than the old treatment the new treatment should be (perhaps by applying decision analysis). Instead of placing the cut-off point at 40.4% as in figure 4, the decision-maker might choose 45% to provide a minimum difference in treatment response rate of 45.0 - 40.4 = 4.6%. The probability of replication within the difference range of over 45% then would be 0.88.

## 11.4 The reproducibility of study methods

These calculations are based on the assumption that the observations in a study can be regarded as a random sample from an infinite number of observations in a population that has a fixed mean or proportion. Each one of these infinite number of observations must therefore be made in exactly the same way. Unless these conditions are met, then it is not appropriate to apply a random sampling model. This inappropriate situation might arise if the study were to be repeated in a different setting where other unknown factors operated so that the 'true' mean or proportion was different and/or also the likelihood probability distribution was wider or narrower. This is analogous to a medical situation when the 'between patient variation' of a diagnostic test result is much greater than the 'within patient variation'. It might also happen because the methods were not described accurately or if some observations were omitted through ignorance, mistakes or dishonesty.

If it is certain that a study was 'perfectly reproducible' by having been conducted impeccably as described so that it were possible to repeat it in exactly the same way in the same setting then the calculated 'idealistic' probability of 0.95 could also be regarded as the 'realistic' probability of replication. However, if it was suspected that there was only a probability of 0.9, for example, that the study methods and subjects were 'perfectly reproducible', the probability of a repeat study result falling within the specified range would be lower than 0.95 so that the latter could be regarded as its approximate upper bound [7]. Others have also suggested previously that a P-value could be regarded as a lower bound for the corresponding posterior probability [8, 9].

**11.5 A 'realistic to idealistic' probability index for different settings**

If the probability of the study methods and subjects being 'perfectly reproducible' were 0.9 and the 'idealistic' probability of replication in some range were 0.95, the probability of repeating the study in exactly the same way and for the result to be in the same replication range would be 0.9*0.95 = 0.855. The probability of the methods and subjects not being 'perfectly reproducible' would be therefore 1-0.9 = 0.1. If it were not 'perfectly reproducible' and the probability of replication was as poor as possible (i.e. 0) then the 'realistic probability of replication in this situation would be 0.1*0 = 0. However if the probability of replication when the study was not 'perfectly reproducible' was as the same as when it was perfectly reproducible (i.e. 0.95) the probability of replication if the study were done differently would be 0.1*0.95 = 0.095. So the 'realistic' probability of replication would be between 0.855+0 and 0.855+ 0.095 (i.e. between 0.855 and 0.95). In other words, the idealistic probability of replication can be regarded as an upper bound for the probability of replication [7] (or lower bound for a P-value [8, 9]) irrespective of the probability of the methods and subjects being perfectly reproducible. It might be reasonable to use the lower bound (e.g. 0.855) as a 'realistic to idealistic' probability (I/R) index.

A reduced 'realistic' probability of replication in other settings (as reflected by an 'I/R index' of less than 1) may be more marked in some disciplines than others because of the presence of more unknown confounding factors affecting the results (e.g. in psychology compared to physics). In their study to estimate replication in psychological science, the Open Science Collaboration [10] discovered that only 47% of effect sizes fell within the 95% confidence intervals of the original studies. If we regard these as an approximate 'idealistic' posterior probability of replication of 0.95 and an approximate 'realistic' posterior probability of replication of 0.47, then the approximate 'I/R index' for this replication range would be 0.47/0.95 = 0.49. If such I/R index were found to be reasonably consistent for a particular discipline or particular type of study and some ranges, it might provide a crude estimate of the probability of replicating a result found in one setting when the study was repeated in other settings. An 'I/R' index would only apply to a specified replication range (e.g. the 95% confidence interval). A difference I/R index might apply for some other range (e.g. a 99% confidence interval). Comparing probability distributions in different settings would be more informative.

**11.6 Comparing probability distributions in different settings**

It may be more informative to assess the difference in the posterior probability distributions in different settings. If there were more uncontrollable confounding factors present in other settings, this would increase the variance and 'broaden' the likelihood distribution. It is also possible that the 'true' mean or proportion might be different in other settings. The Open Science Collaboration [10] discovered that there was an increased in variance of the effect size combined with a diminished actual effect size in repeat studies. Different hypothetical distributions could be used as a form of sensitivity analysis together with power calculations to plan future studies in other settings. Perhaps the practical answer would be to regard the interesting result of a study with a high 'idealistic' probability of replication in one setting as a pilot study to be repeated in a collaborative 'multi-settings' study and ideally with 'registered reports' [11] before the findings became widely accepted.

**11.7 Updating knowledge with new study results**

The results of previous studies are usually described in the introduction of the research report. Some advocate that such results of previous studies should be pooled in a meta-analysis in a research paper's introduction and that the results of the new study should be added to these pooled results to update the shared experience. This can be done in a frequentist manner by using Bayes rule. Bayesians also advocate combining personal experience (as opposed to well-documented past data) with new data in this way. A fresh unbiased study performed meticulously will continue to converge on the true mean as the number of observations increase. However, if a hypothetical Bayesian prior probability distribution does not share the same true mean or proportion, it will bias the fresh study and delay its convergence to a true mean or proportion and may thus be counter-productive.

**11.8 Choosing prior probabilities**

Senn [12] has emphasised the variety of ways in which 'Bayesian' prior probabilities are chosen for statistical and the dubious nature of some of these ways. With this in mind, it is important to distinguish between the types of hypotheses under consideration. It has been shown here that a Bayesian prior probability distribution to be used for statistical inference should be based on postulated data that might be obtained from the current study's methods, perhaps based on experience of similar previous studies. This does not include prior probabilities of population proportions (e.g. the probability distribution of prevalence of bilingualism in various countries around the world does not affect the result of random sampling of a particular population being studied).

**11.9 Frequentist and Bayesian applications of Bayes rule**

Bayes rule can therefore be used in 'frequentist' and 'Bayesian' ways, for example:

1. It can be used in a frequentist manner to calculate a posterior probability from the likelihood distribution of a single data set.

2. It can be used in a frequentist manner to perform a meta-analysis using past and current data by combining their likelihood distributions.
3. It can be used in a 'Bayesian' manner to try to predict what would happen to the probability of replication if various hypothetical new data from different settings were incorporated into a future study
4. It can be used in a 'Bayesian' manner to combine probabilities based on personal impressions with fresh data

## 12. Conclusion

For random selection models, the 'base-rate' prior probabilities of all possible outcomes are equal. This allows a 'frequentist' posterior probability of a 'true' outcome (based on an infinitely large number of observations) falling within any specified range to be calculated by applying Bayes rule to the likelihood distribution generated by study data alone. This would be an 'idealistic' posterior probability of replication within a specified range under perfect sampling conditions and could be regarded as an approximate upper bound to a 'realistic' probability of replication under less than perfect replicating conditions. The frequentist posterior probability of replication may be easier for non-statisticians to understand and to apply than the troubled concept of the P-value.